\documentclass[12pt]{iopart}
\usepackage{iopams}
\usepackage{graphicx}
\usepackage{subfigure}
\usepackage{amssymb}

\begin{document}
\title{Entanglement dynamics of a moving multi-photon Jaynes-Cummings
model in mixed states}

\author{Lei Tan$^*$, Zhong-Hua Zhu, and Yu-Qing Zhang}
\address{Institute of Theoretical Physics, Lanzhou University,
              Lanzhou 730000, China}

\eads{\mailto{$^*$tanlei@lzu.edu.cn}}

\begin{abstract}
Using the algebraic dynamical method, the dynamics of entanglement
in an atom-field bipartite system in a mixed state is
investigated. The atomic center-of-mass motion and the field-mode
structure are also included in this system. We find that the
larger values of the detuning and the average photon number, the
smaller fluctuation of the entanglement, but the period for the
evolution of the entanglement doesn't increase accordingly; It is
also found that the fluctuation of the entanglement varies
slightly with the atomic motion and oscillates fast with the value
of the field-mode structure and the transition photon number
increasing. Moreover, a damping evolution of the entanglement
appears when considering detuning and the atomic motion
simultaneously.
\end{abstract}

\submitto{\PS}

\section{Introduction}

 Entanglement has been widely
investigated in quantum information processing
\cite{Vag,Fic,Vit,Zha,Ye} and plays important roles in many
potential applications, such as quantum communication, quantum
teleportation, quantum cryptography, entanglement swapping, dense
coding, and quantum computing, etc
\cite{Ben,Met,Eke,Wie,Son,Nie,Sho}.
 Moreover, with the recent rapid developments in quantum information,  more precise
quantifications for entanglement are required to produce,
manipulate and detect entanglement states. Most of earlier studies
pay much attention to  the bipartite systems that start in pure
states, which are presently well understood \cite{Ber}. While for
mixed states,  the entanglement is difficult to be measured,
because it is not easy to define an analog of the Schmidt
decomposition. However, it is more significant for studying the
systems in mixed states since there doesn't exist a system that
could be decoupled perfectly from environmental influences for
realistic states observed in experiments. Several entanglement
measures for mixed states have been proposed, such as the
entanglement of formation $E_{F}(\rho)$, the entanglement cost
$E_{C}(\rho)$ and the distillable entanglement $E_{D}(\rho)$
\cite{Div,Pop}, but these are not effective computational means.
To overcome these shortcomings, Peres and Horodecki have proposed
a new criterion for separability \cite{Per,Hor}. Based on this
criterion, negativity is introduced, which is a good quantity for
measuring the entanglement of mixed states \cite{Vid,Vida} and is
equal to absolute value of the sum of negative eigenvalues of the
partial transpose of density operator with a bipartite mixed
state.

There are considerable researches on the dynamics of entanglement
for the Jaynes-Cummings model (JCM) \cite{Jay} with a mixed state.
The entanglement for the model consisting of  a two-level atom
interacting with a single mode field through a two-photon process
was studied \cite{Zho}. In \cite{Oba}, the authors devoted to
studying the effects of phase damping on the entanglement
evolution, where the two-level atom initially in a mixed state
interacts with a  multi-quanta single mode quantized  cavity field
including cavity loss. Akhtarshenas and Farsi \cite{Akh} studied
the entanglement of the JCM with the atom initially in mixed state
and the field prepared in Fock states. Very recently, the
entanglement for the JCM with a two-level atom in a lossy cavity
was manipulated by a classical driving field \cite{Zhan}, which
was a novel scheme to control the behavior of the entanglement. Xu
and co-workers \cite{Xu} studied the entanglement dynamics of a
three-level atom in $\Xi$ configuration, and they found that the
maximal value of entanglement decreases with the temperature of
the thermal field and can be controlled by the detuning. Zhang and
Xu \cite{Zhang} considered the entanglement of JCM interacting
with two mode quantized fields for two different initial
situations. Yon$\ddot{a}$c \cite{Yonac-Yu} et al studied the
entanglement dynamics of two separate Jaynes-Cummings Hamiltonians which consist of two
identical two-level atoms inside two separate cavities,
respectively. They investigated how the two sets of pure initial
states affected the entanglement dynamics with resonant situations.
The  literature \cite{Dur-Bri} has shown that the pure entangled states are very difficult to obtain
due to the interaction of uncontrollable environment and thus the
atomic initial state prepared in mixed states is more practical and
general.
In this paper, we will focus on the problem of a
moving two-level atom coupled with a quantized field in a
multi-photon transition process. The effects of the detuning, the
transition photon number, atomic motion, the field-mode structure
and the temperatures on the entanglement are all discussed. In
fact, the model including all these factors is extremely difficult
to deal with using the conventional methods. Here we adopt the
algebraic dynamical approach \cite{Wan,Jie} to study the nonlinear
system, and the key idea is introducing a canonical transformation
that transforms the Hamiltonian into a liner function in terms of
a set of Lie algebraic generators. The nonlinear system is thus
integrable and solvable, then the density operator with time
evolution can be derived easily.

The paper is organized as follows: In Section 2, we present the
basic model, and derive the time evolution expression of the
density matrix  using the algebraic dynamical approach. In Section
3, we study the dynamics evolution of the entanglement between the
two subsystems for different situations with the atom initially in
a pure state and  a mixed state. Finally, a brief conclusion is
given in Section 4.

\section{The multi-photon JCM}

We investigate a moving two-level atom interacting with a single
mode quantized radiation field in a multi-photon transition
process. Within the rotating wave approximation, the Hamiltonian
of the bipartite system can be described as
\begin{eqnarray}\label{eq:1}
H=\omega_{0}S_{z}+\omega a^{\dagger}a+g[f(z)]^{l}(a^{\dagger
l}S_{-}+a^{l}S_{+})   \quad(\hbar=1),
\end{eqnarray}
where $\omega_{0}$ is the atomic transition frequency, $\omega$ is
the field frequency. $a^{\dagger}$ ($a$) denote the creation
(annihilation) operators of the radiation field and $S_{+}$
($S_{-}$) represent the atomic raising (lowering) operators.
$S_{z}$ is the atomic inversion operator. $g$ is the atom-field
coupling constant. $f(z)$ is the shape function of the cavity
field mode \cite{Sch} and the atomic motion can be incorporated as
\begin{eqnarray}\label{eq:2}
f(z)\rightarrow f(\upsilon t),
\end{eqnarray}
where $\upsilon$ is the atomic motion velocity. A specific mode
$TEM_{mnp}$ is defined:
\begin{eqnarray}\label{eq:3}
f(z)=\sin(\frac{p\pi\upsilon t}{L}),
\end{eqnarray}
with $p$ represents the number of half-wavelengths of the field
mode inside a cavity of the length $L$. When the atomic motion
velocity is considered as $\upsilon=gL/\pi$, the evolution
function $\theta(t)$ can be obtain
\begin{eqnarray}\label{eq:4}
\theta(t)=\int_{0}^{t}[f(\upsilon t')]^{l}dt'.
\end{eqnarray}
We restrict the motion of the atom along the z-axis. There exists
one conservative quantity for the system as follow:
\begin{eqnarray}\label{eq:5}
N=a^{\dagger}a+l(S_{z}+\frac{1}{2}),
\end{eqnarray}
which commutes with Hamiltonian (1). To linearize the Hamiltonian
by the algebraic dynamical approach, a canonical transformation
operator is introduced as
\begin{eqnarray}\label{eq:6}
U_{g}=\exp[\frac{\alpha}{F^{1/2}}(a^{l} S_{+}-a^{\dagger l}
S_{-})],
\end{eqnarray}
where $\alpha=-\arctan[(\sqrt{\Delta^{2}/4+\tilde{g}^{2}
F}-\Delta/2)/ \tilde{g} F^{1/2}]$, $\Delta=\omega_{0}-l\omega$,
$\tilde{g}=g\theta(t)/t$, and $F=N!/(N-l)!$. Based on the
algebraic dynamical method,  SU(2) algebra generators $\{J_{0}$,
$J_{+}$, $J_{-}\}$ are introduced, with
 $J_{0}=S_{z}$,
$J_{+}=F^{-1/2}a^{l}S_{+}$, $J_{-}=F^{-1/2}a^{\dagger l}S_{-}$,
which satisfy the following commutation relations:
\begin{eqnarray}\label{eq:7}
[J_{0},J_{+}]=J_{+},\quad [J_{0},J_{-}]=-J_{-}, \quad
[J_{+},J_{-}]=2J_{0}.
\end{eqnarray}
Then the canonical transformation operator is reduced as
\begin{eqnarray}\label{eq:8}
U_{g}=\exp[\alpha (J_{+}-J_{-})].
\end{eqnarray}
After the canonical transformation, the dressed Hamiltonian can be
written in terms of the SU(2) algebra generators
\begin{eqnarray}\label{eq:9}
H'=U_{g}^{-1}(\int H dt)U_{g}=E(N)+\lambda J_{0},
\end{eqnarray}
where $$\lambda=\sqrt{\Delta^{2}+4\tilde{g}^{2}F}, \quad
E(N)=\omega(N-l/2).$$ Then the time evolution operator can be
expressed by
\begin{eqnarray}\label{eq:10}
U(t)&=&e^{-i\int H dt}\nonumber\\ &=&U_{g}\exp[-i(E(N)+\lambda J_{0})]U_{g}^{-1}\nonumber\\
&=& e^{-iE(N)t}[\cos\frac{\lambda t}{2}-2i J_{0}\sin\frac{\lambda
t}{2}\cos 2\alpha + i(J_{+}+J_{-})\sin\frac{\lambda
t}{2}\sin2\alpha].
\end{eqnarray}

Assuming the cavity field is initially in the single-mode thermal
state
\begin{eqnarray}\label{eq:11}
\rho_{f}(0)=\sum^{\infty}_{n=0}P_{n}|n\rangle\langle n|,\quad
P_{n}=\frac{m^{n}}{(m+1)^{n+1}},
\end{eqnarray}
with $m=1/[\exp(\omega/T)-1]$, is the mean photon number of the
cavity field for thermal equilibrium at a certain temperature $T$,
and the atom is initially in a statistical mixed state
\begin{eqnarray}\label{eq:12}
\rho_{a}(0)=C_{e}|e\rangle\langle e|+C_{g}|g\rangle\langle g|,
\end{eqnarray}
with $C_{e}$+$C_{g}$=1, $C_{g}=0$ ($C_{g}=1$) means the atom is
initially prepared in the excited state (the ground state) and
$0<C_{g}<1$ represents the initial state of the atom is in the
mixed state. Since there is no system that could be decoupled
perfectly from environmental influences and thus mixing is
unavoidable, mixed states are more realistic and meaningful than
pure states in experimental situation \cite{Verstraete}.
  Accordingly, the initial state of the total system
can be given as
\begin{eqnarray}\label{eq:13}
\rho_{af}(0)=\rho_{f}(0)\otimes\rho_{a}(0)
=\sum_{n}P_{n}C_{g}|n,g\rangle\langle
n,g|+\sum_{n}P_{n}C_{e}|n,e\rangle\langle n,e|,
\end{eqnarray}
Using Eqs.(10) and (13), the density operator with the time
evolution can be written by
\begin{eqnarray}\label{eq:14}
\rho_{af}(t)=U(t)\rho_{af}(0)U^{+}(t)
=\sum_{n}P_{n}C_{g}\rho_{g}(t)+\sum_{n}P_{n}C_{e}\rho_{e}(t),
\end{eqnarray}
where
\begin{eqnarray}\label{eq:15}
\rho_{g}(t)&=&[\cos^{2}(\frac{\lambda_{n}t}{2})
+\sin^{2}(\frac{\lambda_{n}t}{2})\cos^{2}(2\alpha_{n})]|n,g\rangle\langle n,g|\nonumber\\
&-&i\sin(\frac{\lambda_{n}t}{2})\sin(2\alpha_{n})[\cos\frac{\lambda_{n}t}{2}
+i\sin\frac{\lambda_{n}t}{2}\cos(2\alpha_{n})]|n,g\rangle\langle n-l,e|\nonumber\\
&+&i\sin(\frac{\lambda_{n}t}{2})\sin(2\alpha_{n})[
\cos\frac{\lambda_{n}t}{2}-i\sin\frac{\lambda_{n}t}{2}\cos(2\alpha_{n})]|n-l,e\rangle\langle
n,g|\nonumber\\&+&\sin^{2}(\frac{\lambda_{n}t}{2})\sin^{2}(2\alpha_{n})|n-l,e\rangle\langle
n-l,e|,
\end{eqnarray}
\begin{eqnarray}\label{eq:16}
\rho_{e}(t)&=&[\cos^{2}(\frac{\lambda_{n+l}t}{2})
+\sin^{2}(\frac{\lambda_{n+l}t}{2})\cos^{2}(2\alpha_{n+l})]|n,e\rangle\langle n,e|\nonumber\\
&-&i\sin(\frac{\lambda_{n+l}t}{2})\sin(2\alpha_{n+l})[\cos\frac{\lambda_{n+l}t}{2}
-i\sin\frac{\lambda_{n+l}t}{2}\cos(2\alpha_{n+l})]|n,e\rangle\langle
n+l,g|
\nonumber\\
&+&i\sin(\frac{\lambda_{n+l}t}{2})\sin(2\alpha_{n+l})[\cos\frac{\lambda_{n+l}t}{2}
+i\sin\frac{\lambda_{n+l}t}{2}\cos(2\alpha_{n+l})]|n+l,g\rangle\langle n,e|\nonumber\\
&+&\sin^{2}(\frac{\lambda_{n+l}t}{2})\sin^{2}(2\alpha_{n+l})|n+l,g\rangle\langle
n+l,g|,
\end{eqnarray}
with $\lambda_{n}=\sqrt{\Delta^{2}+4\tilde{g}^{2}F(n)}$ ,
$\alpha_{n}=-\arctan[(\sqrt{\Delta^{2}/4+\tilde{g}^{2}F(n)}-\Delta/2)/(\tilde{g}F^{1/2}(n))]$,
$F(n)=n!/(n-l)!$.

\section{Entanglement evolution for the bipartite system}

In this section negativity is employed to investigate the
entanglement for the coupled atom-field system. The system is
described by the density matrix in Eq.(14), and the entanglement
degree is quantified by \cite{Vid}
\begin{eqnarray}\label{eq:17}
\mathcal{N}(\rho)\equiv\frac{\|\rho^{T}\|-1}{2},
\end{eqnarray}
where $\rho^{T}$ is the partial transpose of $\rho$.
$\|\rho^{T}\|$ denotes the trace norm of $\rho^{T}$ which is equal
to the sum of the absolute values of all eigenvalues of
$\rho^{T}$. $\mathcal{N}(\rho)$ is equal to the absolute value of
the sum of the negative eigenvalues of $\rho^{T}$, which has been
proposed as a quantitative entanglement measure. It is a good
measure of entanglement because negativity is an entanglement
monotone, does not increase under local operations and classical
communication (LOCC) and can also be  easily treated
mathematically. From the final state in Eq.(14) and using Eq.(17),
the amount of entanglement can be obtained from the following
result
\begin{eqnarray}\label{eq:18}
\|\rho^{T}(t)\|&=&\sum^{l-1}_{n=0}|\xi_{n}|+\frac{1}{2}\sum^{\infty}_{n=0}[
|\mu_{n}+\xi_{n+l}
+\sqrt{(\mu_{n}-\xi_{n+l})^{2}+4\phi_{n}\chi_{n}}
|\nonumber\\&+&|\mu_{n}+\xi_{n+l}
-\sqrt{(\mu_{n}-\xi_{n+l})^{2}+4\phi_{n}\chi_{n}}|],
\end{eqnarray}
where
\begin{eqnarray}\label{eq:19}
\mu_{n}&=&C_{g}P_{n}[\cos^{2}(\frac{\lambda_{n}t}{2})
+\sin^{2}(\frac{\lambda_{n}t}{2})\cos^{2}(2\alpha_{n})]
+C_{e}P_{n-l}\sin^{2}(\frac{\lambda_{n}t}{2})\sin^{2}(2\alpha_{n}),\nonumber\\
\xi_{n}&=&C_{e}P_{n}[\cos^{2}(\frac{\lambda_{n+l}t}{2})
+\sin^{2}(\frac{\lambda_{n+l}t}{2})\cos^{2}(2\alpha_{n+l})]+
C_{g}P_{n+l}\sin^{2}(\frac{\lambda_{n+l}t}{2})\sin^{2}(2\alpha_{n+l}),\nonumber\\
\phi_{n}&=&\chi^{\ast}_{n}\nonumber\\
&=&i(C_{g}P_{n+l}-C_{e}P_{n})\sin(\frac{\lambda_{n+l}t}{2})\sin(2\alpha_{n+l})[
\cos\frac{\lambda_{n+l}t}{2}-i\sin\frac{\lambda_{n+l}t}{2}\cos(2\alpha_{n+l})],
\end{eqnarray}

Based on Eqs.(17) and (18), we present some interesting numerical
results for different parameters to demonstrate the effects on the
entanglement degree. In Fig. 1, the negativity is shown as
functions of $C_{g}$ and the scaled time $gt/\pi$. When the
probability of the atom initially in the ground state is lager,
the entanglement is smaller, and the entanglement is close to zero
as the value of $C_{g}$  exceeds 0.5. Therefore, we choose
$C_{g}=0.2$ to demonstrate the effects of the parameters on the
entanglement. Fig. 2 displays the relation between the
entanglement and the atomic motion and the field-mode structure.
Here, the atomic motion is considered as $\upsilon=gL/\pi$ and is
neglected by performing $\theta(t)\rightarrow t$. It is found that
the atomic motion leads to the regular evolution period of the
negativity by comparing Fig. 2(a) and Fig. 2(b), and the
increasing of $p$ shortens the period of the entanglement and
results in a slight decreasing of the amplitude in Figs. 2(b) and
(c) (In fact, when the parameter varies in a small region, the
amplitude of the entanglement doesn't change a lot).
Interestingly, when the atom motion is considered and the
evolution time is in certain times $gt=2n\pi/p$ $(n=0,1,2\cdots)$,
the negativity goes to zero which is  called ``entanglement sudden
death" (EDS) \cite{Yon,Yan,Schl} and in the other Figs, the
similar phenomenon also occurs.

Now we shall turn our attention to the effects of the detuning.
What is worth noting  that most of previous researches only
consider  the case of the resonance to avoid the complexity.
Without loss of generality, considering the detuning is necessary
and meaningful. In Fig. 3, it clearly shows that when the detuning
increases the entanglement decreases and oscillates fast
\cite{Cum}, and it is interesting to note that when the detuning
and the atomic motion are all considered, a damping evolution of
the entanglement arises. The former is attributed to the fact that
on having large dutunings, the atomic system is weakly coupled to
the radiation field, and the latter may be referred to a quantum
decoherence \cite{Zidan,Tumm,Vagli} when the interaction between
the internal and the external atomic dynamics is considered (see
Fig.3(b)). Also,  it is shown that the atomic motion  can lead to
a damping in the correlation functions of non classical behaviors
between the field and internal atomic variables, which induces the
separability of the atom and the field in \cite{Tumm}. Fig.4
depicts how the average photon number $m$ influences the
entanglement evolution. One can see that with increasing the
average photon number the negativity decreases, but the evolution
periodicity of the entanglement doesn't change. These results show
that the maximum value of the entanglement depends on the mean
photon number of the field (or the thermal equilibrium temperature
$T$), but the period doesn't, which agrees well with the results
in \cite{Shao}. As a matter of fact, for the thermal field, when
the average photon number increases, the weight mixture of number
states denoted by $P_{n}$ becomes small, so the entanglement
feature is washed out greatly. In addition, different photon
transition processes can also affect the entanglement, as is shown
in Fig.5. Intensity of the entanglement increases as the photon
number $l$ increases, meanwhile the oscillation of the negativity
becomes faster and faster \cite{Abdel}.

Experimentally, the realization of the JCM in this paper is not
difficult. We could inject an highly excited Rydberg atom along
the axis of the cavity which moves fast to eliminate no reflection
effect from the cavity, then we could assume that it has a
constant velocity along the cavity \cite{Joshi}. The transition between the
upper and lower levels of the atom may involve $l$ photons by
adjusting the energy separation between the level close to the
energy of $l$ quanta of the electromagnetic field.
Photons being the fastest information carriers have
only very weak coupling to the environment, and are thus the most successful physical
system for the observation of multi-partite entanglement and proof-of-principle
demonstrations of a diversity of quantum information applications so far.
Multi-photon entanglement has become a hot topic recently and we believe
the new approach in this paper
will promote the relate development.

\section{Conclusion}

We have used negativity to study the entanglement for the system
of a moving two-level atom interacting with a quantized single
mode field in the mixed state with a multi-photon process. With
the algebraic dynamical method, its solution is obtained. In our
work, the atomic motion and the field-mode structure, the
detuning,  the transition photon number and the average photon
number of the field are all considered. Our studies show that
increasing of the detuning or the average photon number can lead
to decreasing of the negativity, the entanglement oscillates fast
for lager detunings and EDS appears with certain times. Atomic
motion and the field-mode structure also give rise to many
interesting effects on the entanglement. An increase of the
parameter $p$ can shorten of the evolution periodicity of the
entanglement. Importantly, a damping evolution of the entanglement
happens when the detuning and the atomic motion are considered
together. Furthermore, we demonstrate the fact that a lager photon
number $l$ makes it quick to the oscillation of the negativity and
bring its amplitude up slightly. Our researches might shed light
on understanding the entanglement of the JCM and  provide a simple
means to study dynamics of the entanglement.

\ack
This work was partly supported by the National Natural Science
Foundation of China under Grant No. $10704031$, the National
Natural Science Foundation of China for Fostering Talents in Basic
Research under Grant No.$J0730314$, and the Natural Science
Foundation of Gansu Under Grant No. 3ZS061-A25-035.

\newpage
\begin{figure}
\centering \includegraphics[width=0.60\textwidth]{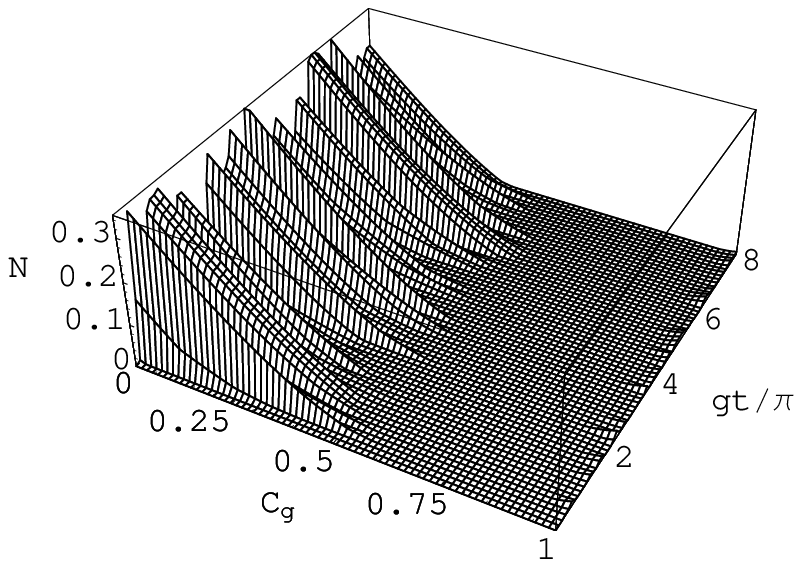}
\caption[fig1]{The evolution of the negativity as  functions of
$C_{g}$ and the scaled time $gt/\pi$. The atomic motion is
considered, other parameters: $m=1$, $\Delta=0$, $p=1$, $l=2$. }
\label{Fig1}
\end{figure}

\newpage
\newpage
\begin{figure}
 \centering
  \subfigure[]{
    \label{fig:subfig:a} 
    \includegraphics[width=2.1in]{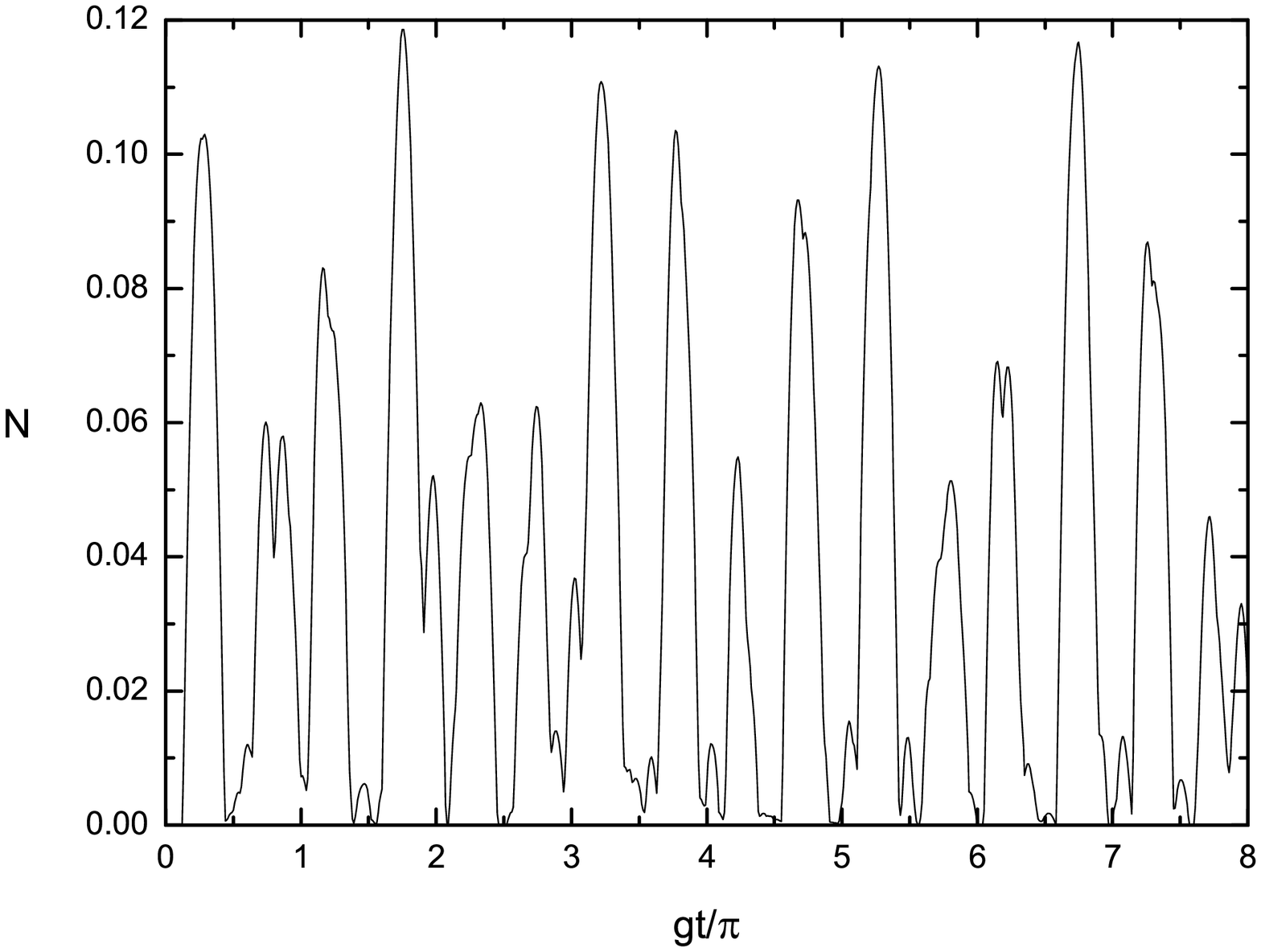}}
  \hspace{1in}
  \subfigure[]{
    \label{fig:subfig:b} 
    \includegraphics[width=2.1in]{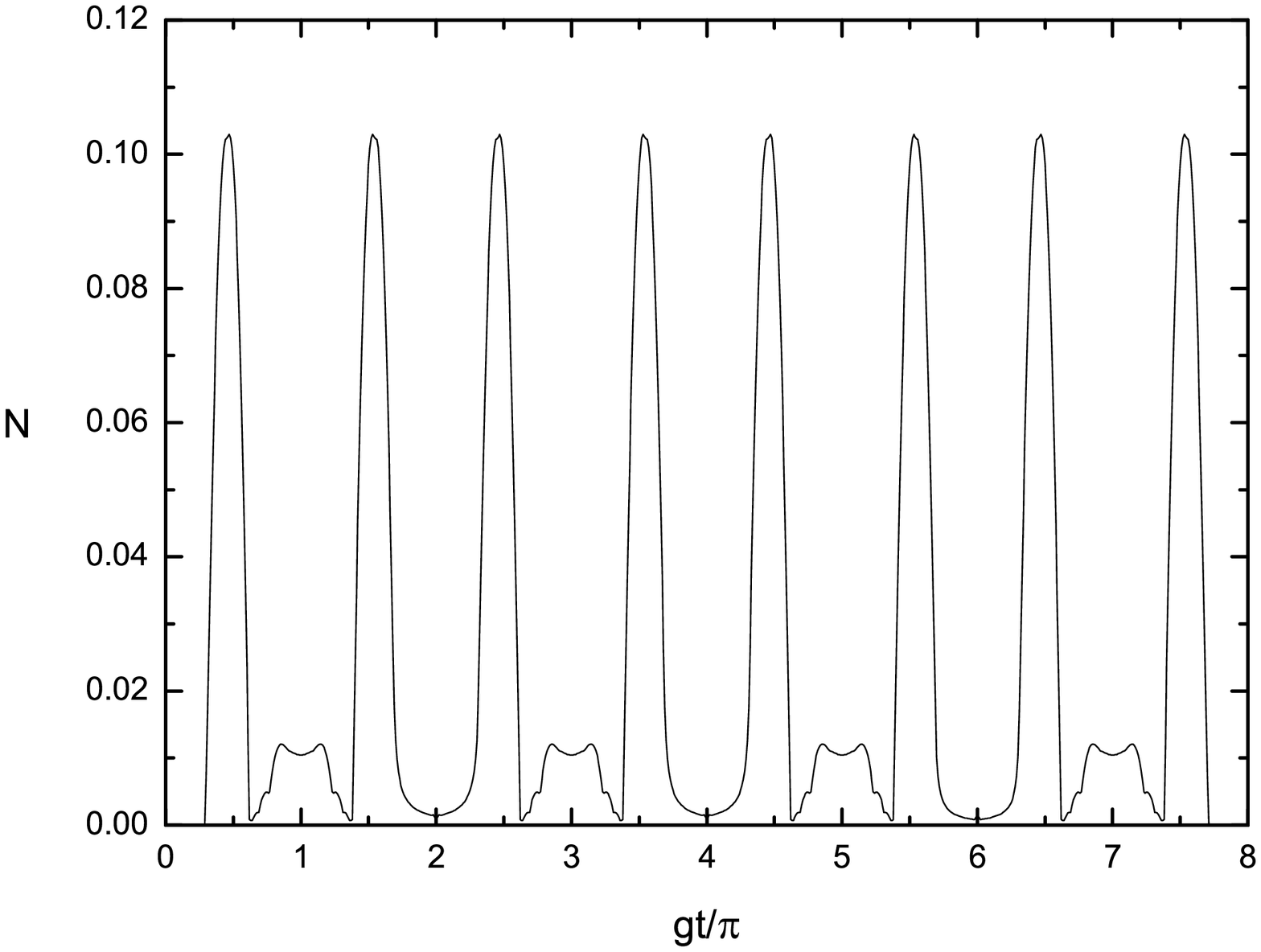}}
  \subfigure[]{
    \label{fig:subfig:b} 
    \includegraphics[width=2.1in]{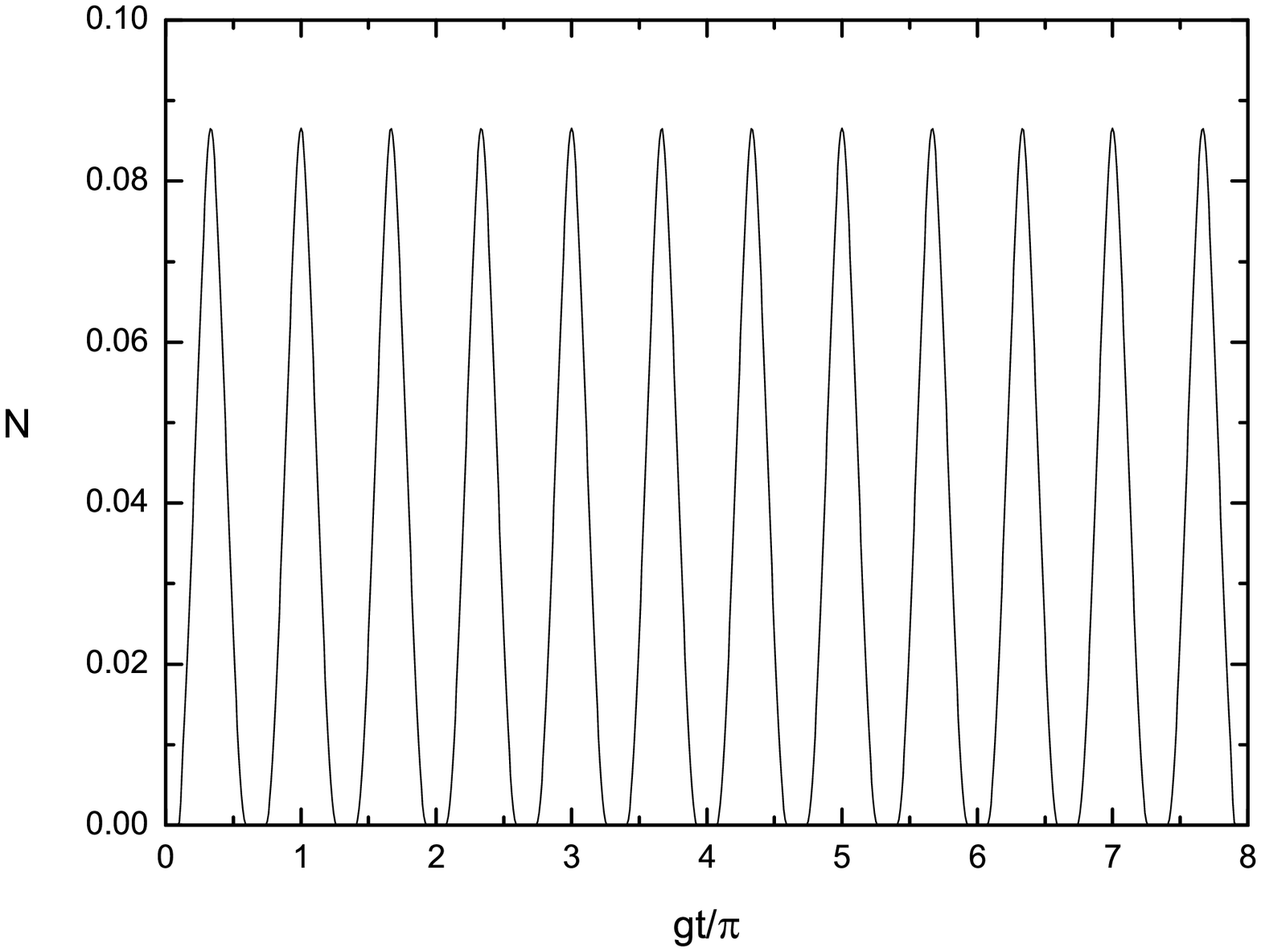}}
  \caption{The evolution of the negativity as a function of the
scaled time $gt/\pi$ for different parameters $p$. Other
parameters: $m=1$, $\Delta=0$, $l=1$, $C_{g}=0.2$,
 (a) the atomic motion is neglected, (b) $p=1$, (c) $p=3$.}
  \label{fig:subfig} 
\end{figure}

\newpage
\newpage
\begin{figure}
 \centering
  \subfigure[]{
    \label{fig:subfig:a} 
    \includegraphics[width=2.1in]{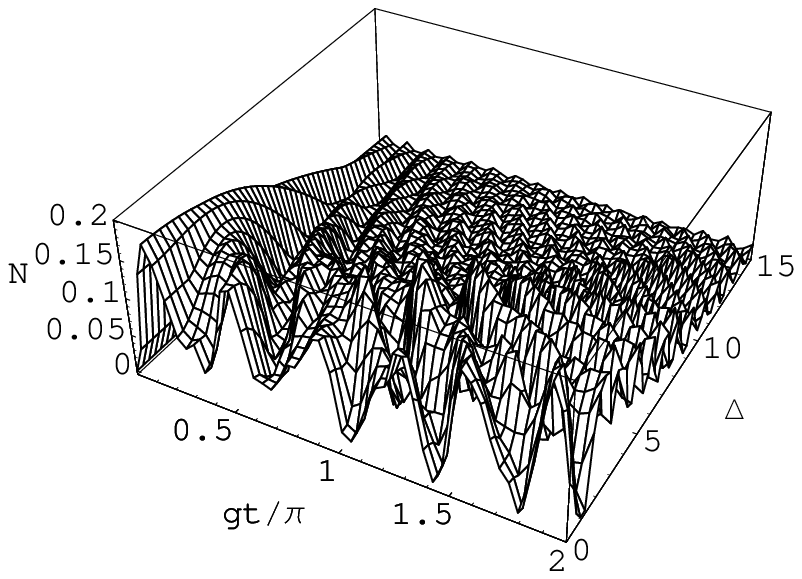}}
  \hspace{1in}
  \subfigure[]{
    \label{fig:subfig:b} 
    \includegraphics[width=2.1in]{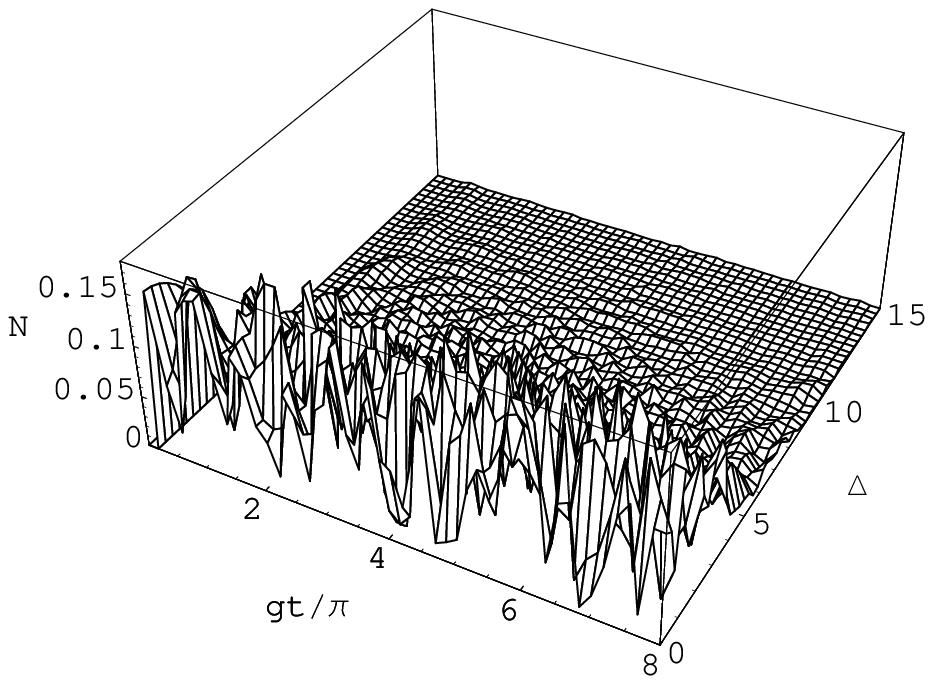}}
   \caption{The evolution of the negativity as functions of the
scaled time $gt/\pi$ and $\Delta$. Other parameters: $m=1$, $l=2$,
$C_{g}=0.2$, (a) the atomic motion is neglected, (b) the atomic
motion is considered ($p=1$).}
  \label{fig:subfig} 
\end{figure}

\newpage
\newpage
\begin{figure}
\centering {      \includegraphics[width=0.60\textwidth]{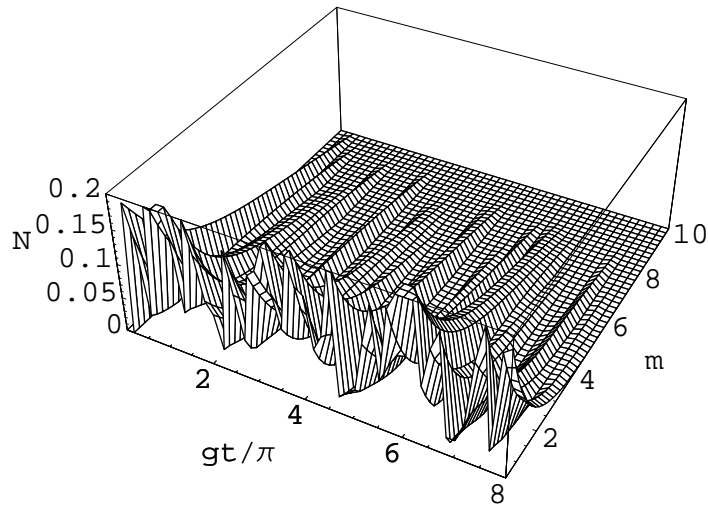}
      \label{fig_firstsub}
} \caption{The evolution of the negativity as functions of the
scaled time $gt/\pi$ and the average photon number $m$. The atomic
motion is considered, other parameters: $\Delta=0$, $p=1$, $l=2$,
$C_{g}=0.2$.} \label{fig_subfigures}
\end{figure}

\newpage
\newpage
\begin{figure}
 \centering
  \subfigure[]{
    \label{fig:subfig:a} 
    \includegraphics[width=2.1in]{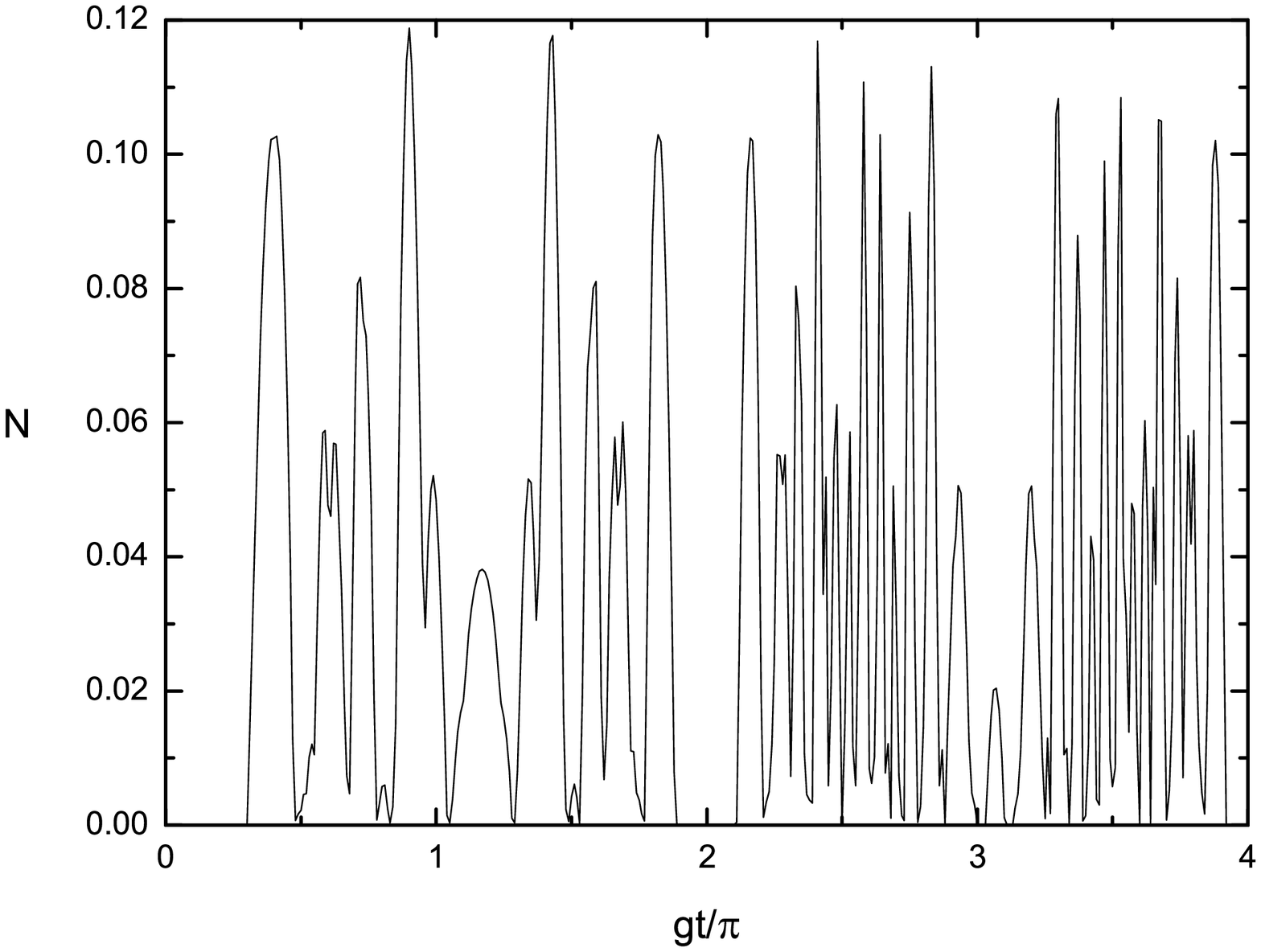}}
  \hspace{1in}
  \subfigure[]{
    \label{fig:subfig:b} 
    \includegraphics[width=2.1in]{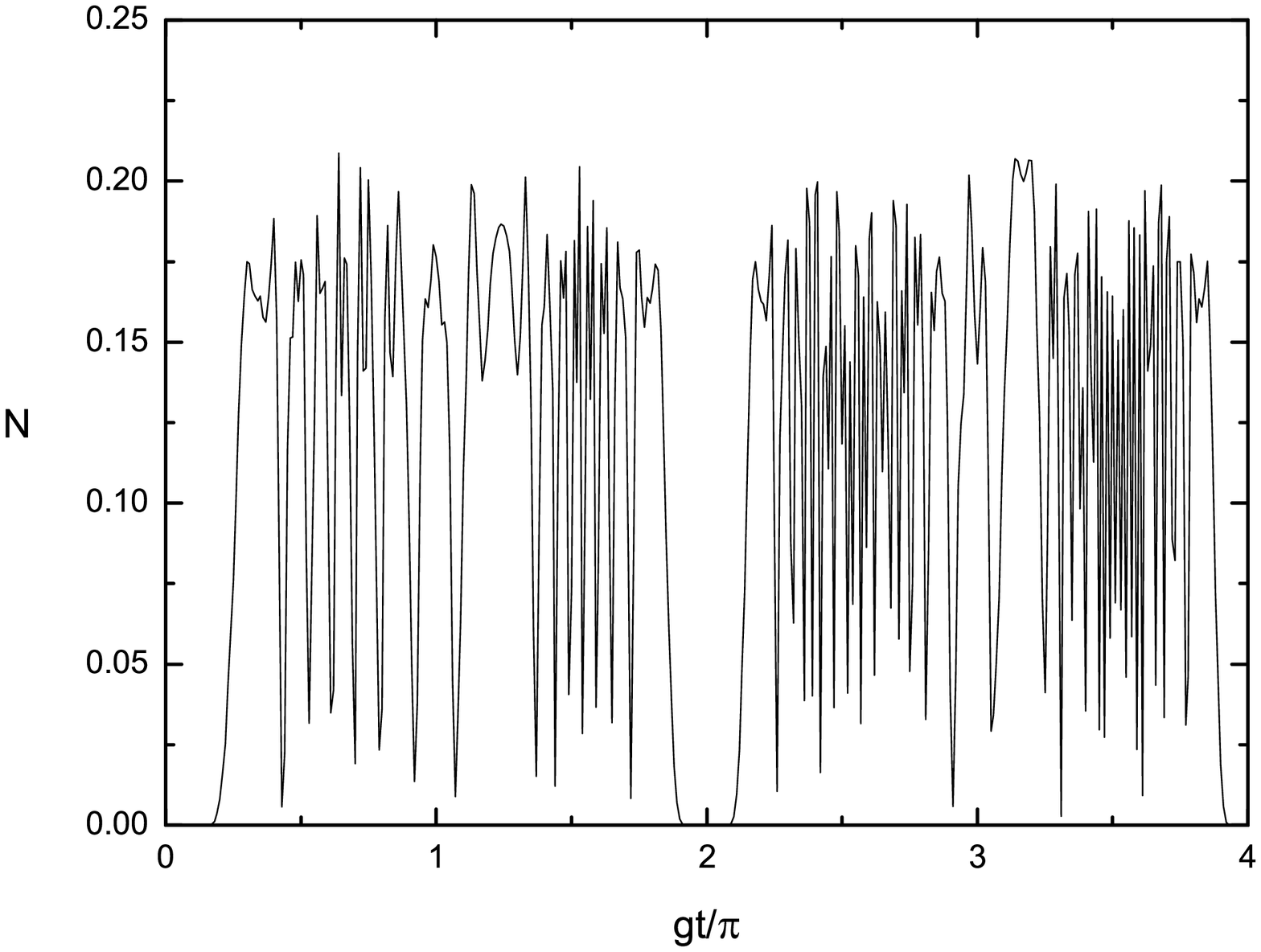}}
  \subfigure[]{
    \label{fig:subfig:b} 
    \includegraphics[width=2.1in]{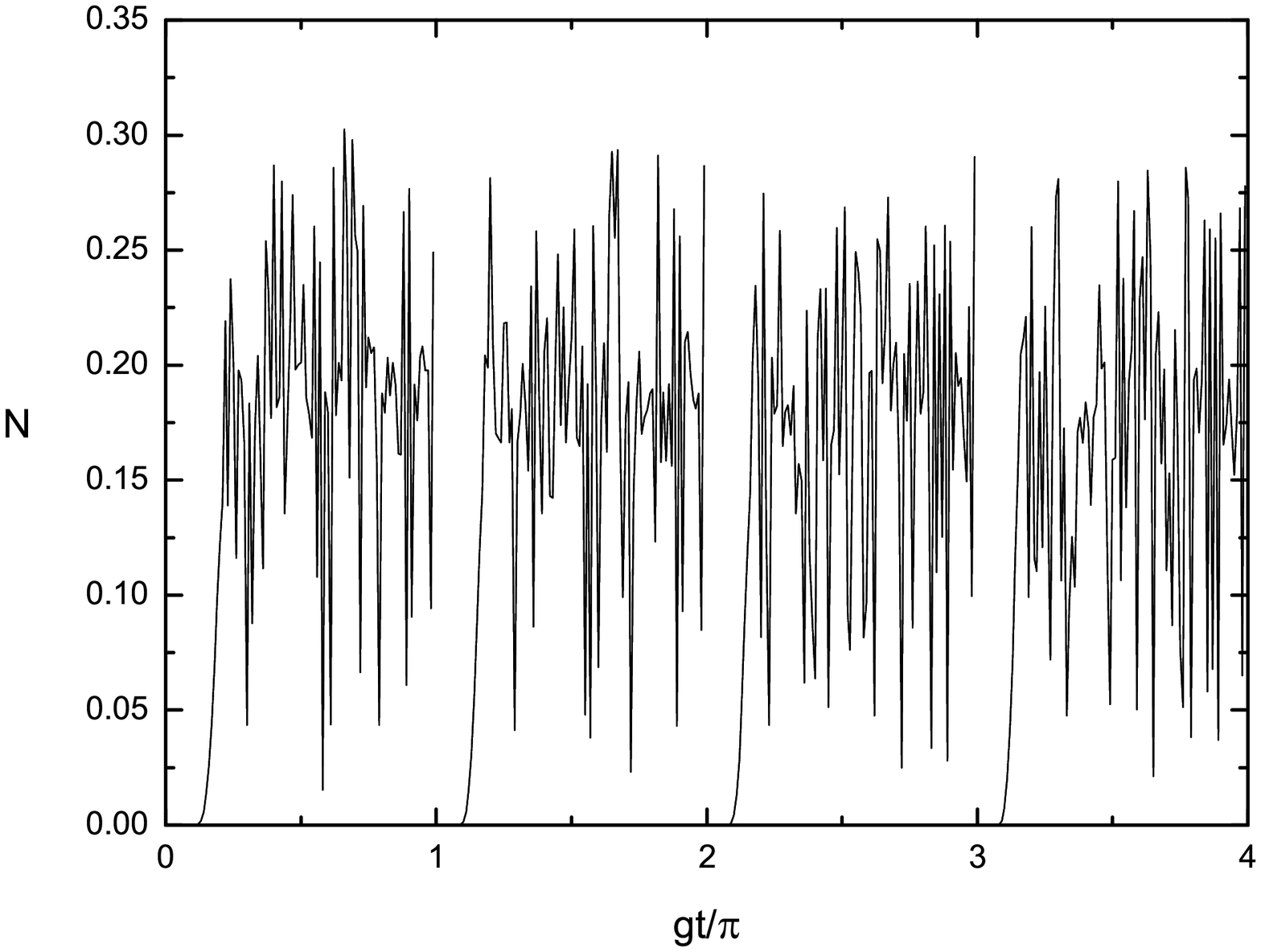}}
  \caption{The evolution of the negativity as a function of the
scaled time $gt/\pi$ for different parameters $l$. The atomic
motion is considered, other parameters: $m=1$, $\Delta=0$, $p=1$,
$C_{g}=0.2$, (a) $l=1$, (b) $l=3$, (c) $l=8$. }
  \label{fig:subfig} 
\end{figure}


\section*{References}
\begin{thebibliography}{99}
\bibitem{Vag}  Vaglica A and  Vetri G 2007 {\it Phys. Rev. A} {\bf 75} 062120


\bibitem{Fic}   Ficek Z and  Tanas R 2006  {\it Phys. Rev. A} {\bf 74} 024304

\bibitem{Vit}   Vitali D,   Gigan S,   Ferreira A,   Bohm H R,   Tombesi P,
  Guerreiro A,   Vedral V,  Zeilinger A and  Aspelmeyer M 2007 {\it Phys. Rev. Lett.}  {\bf 98} 030405

\bibitem{Zha}   Zhang Q and Zhang E Y 2002 {\it Acta Phys.
Sin.} {\bf 51} 1684

\bibitem{Ye}   Ye L and Guo G C 2002
 {\it Chin. Phys.} {\bf 11} 996


\bibitem{Ben}    Bennett C H, Brassard  G,   Cr$\acute{e}$peau C,  Jozsa R,
 Peres A and   Wootters W K 1993
{\it   Phys. Rev. Lett.}  {\bf 70}  1895

\bibitem{Met}
 Metwally  N,   Abdelaty M and   Obada A-S F 2004
{\it   Chaos, Solitons and Fractals} {\bf 22} 529


\bibitem{Eke}   Ekert A K 1991
{\it   Phys. Rev. Lett.}  {\bf 67}  661

\bibitem{Wie}  Bennett C H and  Wiesner S J 1992
{\it   Phys. Rev. Lett.} {\bf 69}  2881

\bibitem{Son}   Yang M,   Song W and Cao Z L  2005
{\it   Phys. Rev. A} {\bf 71}  034312

\bibitem{Nie}   Nielsen M A and  Chuang I L 2000 {\it Quantum computation and
quantum information, Cambridge, England}

\bibitem{Sho}  Bennett C H,   Shor P W,   Smolin J A and Thapliyal A V 1999
 {\it   Phys. Rev. Lett.} {\bf  83}  3081

\bibitem{Ber}  Bennett C H,  Bernstein H J,  Popescu S and Schumacher B 1996
 {\it   Phys. Rev. A} {\bf  53}   2046

\bibitem{Div}  Bennett C H,   Divincenzo D P,  Smolin J A and   Wootters  W K 1996
{\it   Phys. Rev. A} {\bf  54}  3824

\bibitem{Pop}   Popescu S and  Rohrlich D  1997
 {\it   Phys. Rev. A} {\bf 56}  3319(R)

\bibitem{Per}  Peres A  1996
{\it   Phys. Rev. Lett.} {\bf 77}  1413

\bibitem{Hor}   Horodecki M,   Horodecki P and  Horodecki R 1996
 {\it   Phys. Lett. A} {\bf 223}  1

\bibitem{Vid}  Vidal G and Werner  R F  2002
 {\it   Phys. Rev. A}  {\bf 65}   032314

\bibitem{Vida}     Vidal G 2000
{\it  J. Mod. Opt.} {\bf 47} 355

\bibitem{Jay}   Jaynes E T and  Cummings F W  1963
{\it  Proc. IEEE.}  {\bf 51}  89

\bibitem{Zho}   Zhou L,  Song H S and  Li C 2002
 {\it  J. Opt. B} {\bf 4}  425

\bibitem{Oba}  Obada A-S F,  Hessian H A and Mohamed A-B A 2007
  {\it   Opt. Commun.} {\bf 280} 230

\bibitem{Akh}  Akhtarshenas S J and Farsi M 2007
{\it   Phys. Scr.}   {\bf 75}  608

\bibitem{Zhan}   Zhang J S and  Xu J B  2009
{\it    Opt. Commun.} {\bf 282}  2543

\bibitem{Xu}  Xu H S and  Xu J B 2009
{\it   Chin. Phys. Lett.} {\bf 26} 010301

\bibitem{Zhang}  Zhang J S and Xu  J B  2009
 {\it   Chin. Phys. B} {\bf 18}   2288

 \bibitem{Yonac-Yu}   Yon$\ddot{a}$c M,   Yu T and  Eberly J H  2007
{\it  J. Physics. B} {\bf 40} S45

\bibitem{Dur-Bri}
  Dur W and  Briegel H J  2007
{\it  Rep. Prog. Phys.} {\bf 70}  1381

\bibitem{Wan}  Wang S J,   Li F L and   Weiguny A 1993
{\it   Phys. Lett. A} {\bf 180}  189

\bibitem{Jie}   Jie Q L, Wang  S J and Wei L F 1997
 {\it   J. Phys. A} {\bf 30}   6147


\bibitem{Sch}  Schlicher R R 1989
 {\it    Opt. Commun.} {\bf 70} 97

\bibitem{Verstraete}
Verstraete F and Verschelde  H 2003
{\it    Phys. Rev. Lett.} {\bf 90}  097901\\
Liao X P,  Fang M F and Zhou Q P 2006
{\it    Physica A} {\bf 365} 351\\
Lee J Y and Kim M S 2000
{\it    Phys. Rev. Lett.} {\bf 84} 4236


\bibitem{Yon}   Yo$\ddot{n}$ac M,  Yu T and Eberly J H 2006
{\it   J. Phys. B} {\bf 39} S621

\bibitem{Yan}
Yan  X Q 2009
{\it   Chaos, Solitons and Fractals}  {\bf 41} 1645


\bibitem{Schl}  Schlosshauer M,   Hines A P and Milburn G J 2008
{\it   Phys. Rev. A} {\bf 77} 022111

\bibitem{Cum}
  Cummings N I and Hu B L 2008
{\it    Phys. Rev. A} {\bf 77}   053823

 \bibitem{Zidan}
 Zidan  N A,  Abdel-Aty M and  Obada A-S F  2002
{\it    Chaos, Solitons and Fractals} {\bf 13}  1421

\bibitem{Tumm}
 Tumminello M,  Vaglica A and Vetri G 2004
{\it   Euro. Phys. Lett.} {\bf 65}  785

\bibitem{Vagli}
 Vaglica A and Vetri G 2007
{\it  Phys. Rev. A} {\bf 75}  062120


\bibitem{Shao}
 Yan  X Q,  Shao B and Zou J 2008
 {\it  Chaos, Solitons and
Fractals} {\bf 37} 835

\bibitem{Abdel}
 Abdel-Aty M,  Furuichi S and  Obada A-S F 2002
 {\it  J. Opt. B} {\bf 4}  37

\bibitem{Joshi}
Joshi A 2010 {\it Opt. Commun.} {\bf 283}  2166

\end{thebibliography}
\end{document}